\documentstyle[twoside,fleqn,espcrc2]{article}

\title{
Self-Duality and Statistical Systems without Internal Energy
Scaling Terms at Criticality
\thanks{Poster}}

\author{
M.~A.~Yurishchev\\
\medskip
{\sl Vasilsursk Laboratory, Radiophysical Research Institute,
606263 Vasilsursk, Nizhny Novgorod Region, Russia}}

\begin{document}

\begin{abstract}
It is argued that self-duality of one system leads to the
{\it zero\/} finite-size scaling amplitude of the critical
internal energy for all system belonging to the same universality
class.
For such models, we may expect that condition of equality (up to
correction-to-scaling terms) of the internal energies for
systems with different sizes will yield more accurate estimates
for the critical temperature than the scaling equation for the
inverse correlation lengths which is used in the standard
phenomenological renormalization-group approach.
Analytical and numerical evidences confirming the above
conjecture are given for examples of two-dimensional
next-nearest-neighbour and spin-1 Ising lattices.
\end{abstract}

\maketitle


Renormalization-group (RG) theory predicts a general structure
of different physical quantities in the vicinity of a phase
transition (\cite{PHA91} and references therein): regular
part plus scaling terms (resulting from relevant fields) and plus
corrections to scaling (connecting with a presence of irrelevant
operators, nonlinear scaling fields, etc.).
For instance, the inverse correlation length of subsystem
with a characteristic linear size $L$ varies at the bulk phase
transition point as
\begin{equation}
   \label{eq:kappaKc}
   \kappa_L(K_c) = L^{-1}(A_\kappa + aL^{-\omega} + \ldots\ ),
\end{equation}
where $A_\kappa$ is the finite-size scaling (FSS) amplitude
of inverse correlation length; $\omega$ and $a$ are the
correction-to-scaling critical exponent and amplitude,
respectively.
Taking two subsystems with sizes $L$ and $L'$ [i.~e. a pair
$(L, L')$] and neglecting the terms of order $O(L^{-\omega - 1})$
in eq.~(\ref{eq:kappaKc}), one obtains the equation
\begin{equation}
   \label{eq:kappaLL1}
   L\kappa_L(K_c)=L'\kappa_{L'}(K_c),
\end{equation}
which serves for estimating $K_c$ in the phenomenological
RG approach \cite{Nigh76}.

Similarly, the internal energy is written as
\begin{equation}
   \label{eq:uKc}
   u_L(K_c) = u_\infty + L^{-d+1/\nu}(A_u + bL^{-\omega}
   + \ldots\ ),
\end{equation}
where $d$ is the space dimensionality and $\nu$ is the
correlation-length critical exponent.

For the derivative of inverse correlation length
$\dot\kappa_L\equiv\partial\kappa_L/\partial K$ we have
\begin{equation}
   \label{eq:dkappaKc}
   \dot\kappa_L(K_c) = L^{-1+1/\nu}(A_{\dot\kappa}
   + cL^{-\omega} + \ldots\ ).
\end{equation}
From this expansion with an accuracy up to terms of order
$O(L^{-\omega-1+1/\nu})$, follows the equation
\begin{equation}
   \label{eq:dkappaLL1}
   L^{1-1/\nu}\,\dot\kappa_L(K_c)
   =(L')^{1-1/\nu}\,\dot\kappa_{L'}(K_c).
\end{equation}

As shown by Privman and Fisher \cite{PF84}, certain
ratios of critical FSS amplitudes are universal quantities.
In particular, the ratio $A_u/A_{\dot\kappa}$ is universal.

Self-duality relation (itself or in combination with the
star-triangle transformation) connects between themselves
the values of a partition function by high and low
temperatures \cite{Bax82,BS96}.
For example, for Ising model on the isotropic square lattice
in the form of $L\times\infty$ strip with periodic boundary
conditions in the transverse direction one has
\begin{equation}
   \label{eq:L1L}
   \lambda_1^{(L)}(K^{*}) = (\sinh2K)^{-L}\lambda_1^{(L)}(K).
\end{equation}
Here $\lambda_1^{(L)}$ is the largest eigenvalue of a transfer
matrix; note that $\lambda_1^{(L)}$ equals the partition function
of a strip per slice.
The dually conjugated value $K^{*}$ is connected with the original
value $K$ by the relation
\begin{equation}
   \label{eq:shKK}
   \sinh 2K^{*}\sinh 2K = 1.
\end{equation}
Since the internal energy is a derivative of the free energy
($u_L=\partial f_L/\partial K$) and the free energy per site
is equal to $f_L=L^{-1}\ln\lambda_1^{(L)}$, then one obtains
from eqs.~(\ref{eq:L1L}) and (\ref{eq:shKK}) that at the critical
point $K^{*}=K=K_c$
\begin{equation}
   \label{eq:uLu8}
   u_L(K_c) = u_\infty .
\end{equation}
That is, due to self-duality of Ising model on a strip, both
the critical FSS amplitude and all correction-to-FSS ones are
zero ($A_u=0$ and all $a=0$).
This is valid also for other isotropic statistical systems which
are invariant under the duality transformations \cite{YuICTP}.

As $A_u/A_{\dot\kappa}$ is the universal quantity, the equality
of the amplitude $A_u$ to zero automatically means the absence
of the leading finite-size term in the internal energy for all
systems (of the same form and with the same boundary conditions)
entering to the universality class of the self-dual system.
In such a case, neglecting the terms of order
$O(L^{-\omega-d+1/\nu})$ in the expansion (\ref{eq:uKc})
we come to the equation
\begin{equation}
   \label{eq:uLuL1}
   u_L(K_c) = u_{L'}(K_c).
\end{equation}
One can expect that this equation will yield estimates of $K_c$
with high accuracy for different systems in the universality
class of which there is a self-dual model.

Using the exact solution for the two-dimensional square Ising
lattice \cite{O44K49} one can establish that in the isotropic
case the dependent-$L$ part is absolutely absent not only in
the internal energy (what was shown above from duality relation)
but also in the temperature derivative of the inverse correlation
length:
\begin{equation}
   \label{eq:dkLKc}
   \dot\kappa_L(K_c) = const\quad {\rm upon}\quad L.
\end{equation}
Therefore the finite-size equation
\begin{equation}
   \label{eq:dkLdkL1}
   \dot\kappa_L(K_c) = \dot\kappa_{L'}(K_c)
\end{equation}
yields the exact value of $K_c$ for the infinite square Ising
lattice from solutions for the strips $L\times\infty$ of finite
widths starting with the pair $(1,2)$.
Moreover, the requirement of agreement of eq.~(\ref{eq:dkLdkL1})
with the scaling relation (\ref{eq:dkappaLL1}) leads \cite{Yu00}
in this model to the exact value for the correlation-length
critical exponent: $\nu=1$.
Therefore, in the Ising model under discussion
\begin{equation}
   \label{eq:dkKc2DI}
   \dot\kappa_L(K_c) = A_{\dot\kappa}\cdot L^0
\end{equation}
with $A_{\dot\kappa}=-2$.
Consequently, in this model all corrections to FSS behaviour in
the inverse correlation length derivative $\dot\kappa_L(K_c)$
are completely compressed.

Let us discuss now the key question, namely the convergence
rate of estimates $K_c$ which follow from
eqs.\~(\ref{eq:uLuL1}) and (\ref{eq:dkLdkL1}).
Take, for instance, models belonging to the two-dimensional
Ising universality class.

At first consider the NNN model, i.~e. a square Ising
lattice with couplings both nearest neighbours (interaction
constant is $J_{NN}$) and next-nearest neighbours (interaction
constant is $J_{NNN}$).
Estimates of $K_c$ for this model are done in \cite{NB82} within
the framework of phenomenological RG approximation with using the
eq.~(\ref{eq:kappaLL1}).
For the double ($L=2$) and triple ($L=3$) Ising strips with $NN$
and $NNN$ couplings, solutions exist in an exact analytical form
\cite{Yu7884}.
Therefore, we used analytical formulas for the internal energy
densities and derivatives of inverse correlation lengths in the
case of a pair $(2,3)$.
Unfortunately, we should restrict ourselves only by numerical
calculations for the strips of larger widths $L$.
The results obtained are collected in table~\ref{tab:NNN1}
($J_{NNN}/J_{NN}=1$) and in table~\ref{tab:NNN14}
($J_{NNN}/J_{NN}=-1/4$).

\begin{table}
\caption{
Estimates of $K_c$ for the 2D NNN Ising lattice by
$J_{NNN}/J_{NN}=1$}
\label{tab:NNN1}
\begin{tabular}{clll}
\hline\\[-3mm]
$(L,L+1)$&${\rm eq.}(2)$&${\rm eq.}(9)$&${\rm eq.}(11)$\\[2mm]
\hline
 $(2,3)$ &0.195084 &0.191796 &0.194076\\
 $(3,4)$ &0.192511 &0.191193 &\\
 $(4,5)$ &0.191374 &0.190596 &\\
 $(5,6)$ &0.190883 &0.190410 &\\
 $(6,7)$ &0.190628 &0.190217 &\\
 $(7,8)$ &0.190484 &0.190269 &\\
 $(8,9)$ &0.190397 &0.190243 &\\[1mm]
 $\infty$&0.19019269(5)  \\
\hline
\end{tabular}
\end{table}

\begin{table}
\caption{
Estimates of $K_c$ for the 2D NNN Ising lattice by
$J_{NNN}/J_{NN}=-1/4$}
\label{tab:NNN14}
\begin{tabular}{clll}
\hline\\[-3mm]
$(L,L+1)$&${\rm eq.}(2)$&${\rm eq.}(9)$&${\rm eq.}(11)$\\[2mm]
\hline
 $(2,3)$ &0.628082 &0.687193 &0.684975\\
 $(3,4)$ &0.643618 &0.694006 &\\
 $(4,5)$ &0.663507 &0.696592 &\\
 $(5,6)$ &0.677302 &0.697329 &\\
 $(6,7)$ &0.685553 &0.697448 &\\
 $(7,8)$ &0.690221 &0.697411 &\\
 $(8,9)$ &0.692834 &0.697357 &\\[1mm]
 $\infty$&0.697\,220(5) & & \\
\hline
\end{tabular}
\end{table}

From these tables it is seen  that approximate values of $K_c$
yielding by eqs.~(\ref{eq:kappaLL1}), (\ref{eq:uLuL1}) and
(\ref{eq:dkLdkL1}) are upper estimates by $J_{NNN}/J_{NN}=1$
while by $J_{NNN}/J_{NN}=-1/4$ those are lower ones.
By this, the accuracy of estimates following from
eqs.~(\ref{eq:uLuL1}) and (\ref{eq:dkLdkL1}) is higher than
that given by eq.~(\ref{eq:kappaLL1}).
In turn, from the two best approximations the estimates of
$K_c$ which are ensured from the internal energy equation have
the highest accuracy.
It is important that the exactness of those estimates by
fixed sizes of subsystems in a pair $(L,L+1)$ is higher than
that which gives eq.~(\ref{eq:kappaLL1}) by bigger subsystem
sizes, i.~e. by larger transfer-matrix orders.
For instance, by $J_{NNN}/J_{NN}=1$ the estimate following from
eq.~(\ref{eq:uLuL1}) with a pair $(2,3)$ is better in comparison
with the estimate which yields eq.~(\ref{eq:kappaLL1}) with the
larger pair $(3,4)$.
For the next pair $(4,5)$ the improvement is much larger.

The second model which we will discuss in the given report is
the square Ising lattice with NN couplings only but with the
spin $S=1$ (in the previous model the spin was equal to
$S={1\over2}$).
Estimates of $K_c$ obtained for such a model within the ordinary
phenomenological RG, i.~e. from eq.~(\ref{eq:kappaLL1}), are
available in \cite{BN85}.
In table~\ref{tab:S1} we reproduce these estimates together
with those which we obtained by solving eq.~(\ref{eq:uLuL1}).
From the table one can see that estimates are lower in
both cases.
By this, higher accuracy is given again by the equation
identifying the critical internal energies of subsystems with
different linear sizes.

\begin{table}
\caption{
Estimates of $K_c$ for the 2D spin-1 Ising lattice}
\label{tab:S1}
\begin{tabular}{clll}
\hline\\[-3mm]
$(L,L+1)$&${\rm eq.}(2)$&${\rm eq.}(9)$\\[2mm]
\hline
 $(2,3)$ &0.579758 &0.589102 \\
 $(3,4)$ &0.584098 &0.588899 \\
 $(4,5)$ &0.587320 &0.589666 \\
 $(5,6)$ &0.588830 &0.590076 \\[1mm]
 $\ \ \ \infty$, [12] &0.5904727(10) \\
\hline
\end{tabular}
\end{table}

Thus, using in fact only the qualitative information that
in the given universality class there is a self-dual system,
we can choose more effective strategy by utilizing the
subsystem solutions which always are restricted by some
maximal size $L_{max}$.

This work was supported by the RFBR through grant 99-02-16472
and by the CRDF through grant RP1-2254.



\end{document}